\documentclass[aps,twocolumn,pra,superscriptaddress,notitlepage,floatfix]{revtex4-2}
\pdfoutput=1

\usepackage{graphicx,color,xcolor}
\usepackage{marvosym}
\usepackage{amsmath,amssymb}
\usepackage{ulem}

\usepackage{graphicx,amsmath,graphics}
\usepackage{amssymb,exscale}
\usepackage{mmap} 
\usepackage[colorinlistoftodos]{todonotes}
\usepackage[colorlinks=true, allcolors=blue]{hyperref}

\newcommand{\ie}{i.\,e.}%

\makeatletter
\def\subsubsection{\@startsection{subsubsection}{3}{10pt}{-1.25ex plus -1ex minus -.1ex}{0ex plus 0ex}{\normalsize\bf}}
\def\paragraph{\@startsection{paragraph}{4}{10pt}{-1.25ex plus -1ex minus -.1ex}{0ex plus 0ex}{\normalsize\textit}}
\renewcommand\@biblabel[1]{#1}
\renewcommand\@makefntext[1]%
{\noindent\makebox[0pt][r]{\@thefnmark\,}#1}
\DeclareRobustCommand\onlinecite{\@onlinecite}
\def\@onlinecite#1{\begingroup\let\@cite\NAT@citenum\citealp{#1}\endgroup}
\def\tagform@#1{\maketag@@@{\ignorespaces#1\unskip\@@italiccorr}}
\let\orgtheequation\theequation
\def\theequation{(\orgtheequation)}
\makeatother

\newcommand{\ry}{Rydberg }

\begin{document}

\title{Ultralong-range Cs-RbCs \ry molecules: non-adiabaticity of  dipole moments }
\author{David Mellado-Alcedo}
 \affiliation{Departamento de F\'{\i}sica, Universidad de C\'ordoba, 14071 C\'ordoba, Spain}
\affiliation{Departamento de F\'{\i}sica At\'omica, Molecular y Nuclear,
 Universidad de Granada, 18071 Granada, Spain}
\author{Alexander Guttridge}
\affiliation{Department of Physics, Durham University, South Road, Durham, DH1 3LE, United Kingdom}

\author{Simon L. Cornish}
\affiliation{Department of Physics, Durham University, South Road, Durham, DH1 3LE, United Kingdom} 
\author{H. R. Sadeghpour}
\affiliation{ITAMP, Center for Astrophysics $|$ Harvard \& Smithsonian , Cambridge, Massachusetts 02138, USA} 
 \author{Rosario Gonz\'alez-F\'erez}
\email{rogonzal@ugr.es}%
\affiliation{Departamento de F\'{\i}sica At\'omica, Molecular y Nuclear,
 Universidad de Granada, 18071 Granada, Spain}
\affiliation{Instituto Carlos I de F\'{\i}sica Te\'orica y Computacional,
 Universidad de Granada, 18071 Granada, Spain} 

\date{\today}

\begin{abstract} 
We consider ultralong-range polyatomic Rydberg molecules formed by combining a \ry cesium atom and a ground-state RbCs molecule. We explore the regime where the charge-dipole interaction due to the Rydberg 
electron with the diatomic polar molecule couples the quantum defect Rydberg states 
Cs($ns$) to the nearest degenerate hydrogenic manifold. 
We consider polyatomic Rydberg molecules in states which are amenable to production in optical tweezers and study the influence of nonadiabatic coupling on the likelihood of their formation.
The decay rates of the vibrational states reflect the interference signature of wave function spread in different coupled potential wells. 
\end{abstract}


\maketitle

\section{Introduction}
\label{sec:introduction}

The exaggerated properties of Rydberg atoms, such as their large spatial extents, high polarizabilities, and energies, make them an outstanding system for precisely probing interactions with the surrounding environment~\cite{Adams_2020}.
In an ultracold setting, a Rydberg atom,  can couple to a ground state atom via the zero-range interaction due to Rydberg electron
low-energy 
scattering from the ground state atom, or interact with another Rydberg atom via van der Waals or dipole-dipole 
forces~\cite{Shaffer2018}. A Rydberg atom can also couple to the permanent electric dipole moment of a polar molecule, via the
electron-dipole interaction~\cite{rittenhouse10,gonzalez15}. 
For dipole moments below 
the Fermi-Teller critical dipole~\cite{fermi47,turner1977}, 
the electron-dipole interaction can produce a molecular bond capable of supporting bound rovibrational states. Such states are predicted to exist in a hybrid atom-molecule gas at ultracold temperatures. 
The prerequisite charge-dipole energy shift was recently observed for the first time in an ultracold hybrid system in an experiment with Rb Rydberg atoms and RbCs molecules trapped in separate optical tweezers \cite{guttridge2023}.

In \ry atoms, core-electron scattering and polarization effects are significant for 
low angular momentum electronic states. Such states are energetically
shifted from the degenerate manifold by means of the so-called quantum defects, which
are the fingerprints that uniquely identify a \ry atom.
The $ns$-states of cesium atom possess a peculiar quantum defect, such that Cs($ns$) states lie 
energetically near the corresponding hydrogenic Cs$(n-4,l\geq 4)$ states, which are nearly degenerate.
Based on this proximity, 
trilobite Rydberg molecules Cs$^*$-Cs possessing sizable ($\sim 1$ kD) permanent electric dipole moments, were observed for the spherically symmetric 
non-degenerate states Cs($ns$)~\cite{tallant12,booth15,peper2021}.
Similarly, when one considers the interaction of a Cs Rydberg atom with a polar molecule, the Cs($ns$) quantum defect facilitates 
hybridization of the \ry states Cs($ns$) and Cs$(n-4,l\geq 4)$, 
which is mediated by the charge-dipole interaction~\cite{GonzalezFerez2020}. 
As a consequence, the optimal scheme for the realization of the ultralong-range triatomic \ry 
molecules~\cite{GonzalezFerez2020}, based on this resonant coupling, also benefits 
from the fact that the diatomic molecule is in the rotational ground state. 

Here, we provide a thorough theoretical study of the shift and formation of the triatomic molecule (TriMol) Cs($ns$)-RbCs.
We first review the basic theory of the ultralong-range triatomic \ry molecules in~\autoref{sec:theory_description}. 
In~\autoref{sec:results}, we consider two
experimental protocols, corresponding to holding the atom and molecule in the same or separate optical tweezers, 
and present the results for selected potential energy curves (PEC) of this TriMol
and their vibrational bound states.
We analyze the impact on the energy shifts of the scattering of the low-energy \ry electron from RbCs.
Inspired by previous studies on  non-adiabatic effects on \ry systems~\cite{peter2023,Duspayev_2022}, we
also explore the role played by the avoided crossings, and provide the Franck-Condon factors for excitation 
from the atomic ground state. The paper is concluded in~\autoref{sec:conclusions}.

\section{The Born-Oppenheimer Hamiltonian}
\label{sec:theory_description}
We consider a TriMol formed by a Rydberg atom and 
a heteronuclear diatomic molecule in the electronic and rovibrational ground state.
The ground state diatomic molecule is described within the Born-Oppenheimer (BO) and 
rigid rotor approximations, whereas the \ry atom is described as a single-electron system.
In this framework, the adiabatic Hamiltonian of the \ry TriMol is 
\begin{equation}
\label{eq:Hamil_adiabatic}
H=H_{\text{A}}+H_{\text{mol}}+ H_{\text{int}}\,,
\end{equation}
where $H_{\text{A}}$ represents the single-electron Hamiltonian describing the \ry atom 
\begin{equation}
\label{eq:Hamil_atom}
H_{\text{A}}=-\frac{\hbar^2}{2m_e}\nabla^2_{r}+V_l(r),
\end{equation}
where $V_l(r)$ is an $l$-dependent nonlocal model potential~\cite{marinescu94}, with $l$ the angular momentum of the \ry electron.
The rigid rotor Hamiltonian describing the molecule is $H_{\text{mol}}=B\mathbf{N}^2 $
with $B$ the rotational constant, and $\mathbf{N}$ the rotational angular momentum.

In this ultracold hybrid system, where the interactions are long-range, the Hamiltonian between the Rydberg atom and the molecule becomes
\begin{equation}
\label{eq:Hamil_molecule}
H_{\text{int}}=-\mathbf{d}\cdot\mathbf{F}_{\text{ryd}}(\mathbf{R},\mathbf{r}) +2\pi a_S(k) \delta\left({\bf r}- {\bf R}\right) ,
\end{equation}
where the first term represents the interaction of the \ry electron and ionic core with the RbCs permanent electric dipole moment 
$\mathbf{d}$. 
The electric field $\mathbf{F}_{\text{ryd}}(\mathbf{R},\mathbf{r})$ due to the Rydberg electron and the ion at the position of the diatomic molecule, $\mathbf{R}$, is  
\begin{equation}
\label{eq:field_rydberg_e_core}
\mathbf{F}_{\text{ryd}}(\mathbf{R},\mathbf{r})=\frac{e}{4\pi\epsilon_0}\left(\frac{\mathbf{R}}{R^3}+\frac{\mathbf{r}-\mathbf{R}}{|\mathbf{r}-\mathbf{R}|^3}\right),
\end{equation}
where $e$ is the electron charge and $\mathbf{r}$ is the position of the \ry electron.
Note that the center of the coordinate system is located at the position of the ionic core, Cs$^+$. 
In this work, we consider the RbCs molecule with a rotational constant of $B=0.490$~GHz~\cite{Gregory2016}
in the electronic and vibrational ground state with an electric dipole moment of 
${d}=1.225$~D~\cite{Molony2014}, 
below the Fermi-Teller critical dipole, $d_{cr} = 1.639$~D~\cite{fermi47,turner1977}. {The subcritical dipole moment of RbCs ensures that the electron does not bind to the molecule.}

The second term in~\autoref{eq:Hamil_molecule} represents the scattering of the 
low-energy, nearly-free electron from the diatomic molecule. This is approximated by the Fermi pseudopotential~\cite{fermi1934,AMOHandbook2023}, assuming that only the $L=0$ ($S$-wave) scattering 
partial wave contributes to the collision.
The $S$-wave scattering length, $a_S(k)=-\tan(\delta_S(k))/{k}$, with $\delta_S(k)$ as the $L=0$ scattering phase shift,
and $k$ the electron wave number, is approximated by a constant value $a_S=1/\sqrt{2E_A}$~\cite{FANO198661}, 
where $E_A$ is the electron affinity of RbCs.

For the numerical description of this system, we fix the center of the laboratory fixed frame at the ionic core Cs$^+$, and 
the diatomic molecule is situated along the $Z$ axis.  
The Schr\"odinger equation associated with the Hamiltonian~\ref{eq:Hamil_adiabatic} is solved for several values of  the interspecies separation $R$.
Hence, the potential energy surfaces of the \ry molecule Cs-RbCs reduce to curves,  
 which only depend on $R$.
To solve this Schr\"odinger equation,
we perform a basis set expansion in terms of the coupled molecular basis~\cite{gonzalez15}
\begin{equation}
\Psi(\mathbf{r},\Omega;R)=\sum_{n,l,N,J} C_{n,l,N}^J(R)\Psi_{nl,N}^{JM_J}(\mathbf{r},\Omega),
\label{eq:basis_expansion}
\end{equation}
where $|l-N|\le J\le l+N$, and 
\begin{equation}
 \Psi_{nl,N}^{JM_J}(\mathbf{r},\Omega)=
\sum_{m_l,M_N}
\langle l m_l NM_N| J M_J\rangle
Y_{NM_N}(\Omega)\,
\psi_{nlm_l}(\mathbf{r}) 
\label{eq:coupled_basis}
\end{equation}
with $\langle l m_l N M_N| J M_J\rangle$  the Clebsch-Gordan coefficients,
$J=|l-N|,\dots,l+N$, and $M_J=-J,\dots,J$. 
$\psi_{nlm_l}(\mathbf{r})$ is the \ry electron wave function
with $n$, $l$ and $m_l$ the principal, orbital and magnetic quantum numbers, respectively, and 
$Y_{NM_{N}}(\Omega)$ the field-free rotational wave function of the diatomic molecule,
whose rotation is described with the Euler angles $\Omega=(\theta, \phi)$.
The total angular momentum of the \ry molecule, excluding an overall rotation, is given by
 $\mathbf{J}=\mathbf{l}+\mathbf{N}$,
where $\mathbf{l}$  is the orbital angular momentum of the electron 
and $\mathbf{N}$ the rotational angular momentum of the diatomic molecule.
We include rotational excitations in RbCs($N\le 5$), the Cs($ns,np,nd,nf$) quantum defects, 
the degenerate manifold, and investigate electronic states of the TriMol with $M_J=0$.

\section{Results}
\label{sec:results}

\subsection{Zero-range electron-molecule scattering}
 \begin{figure}
 \includegraphics[width=0.94\linewidth]{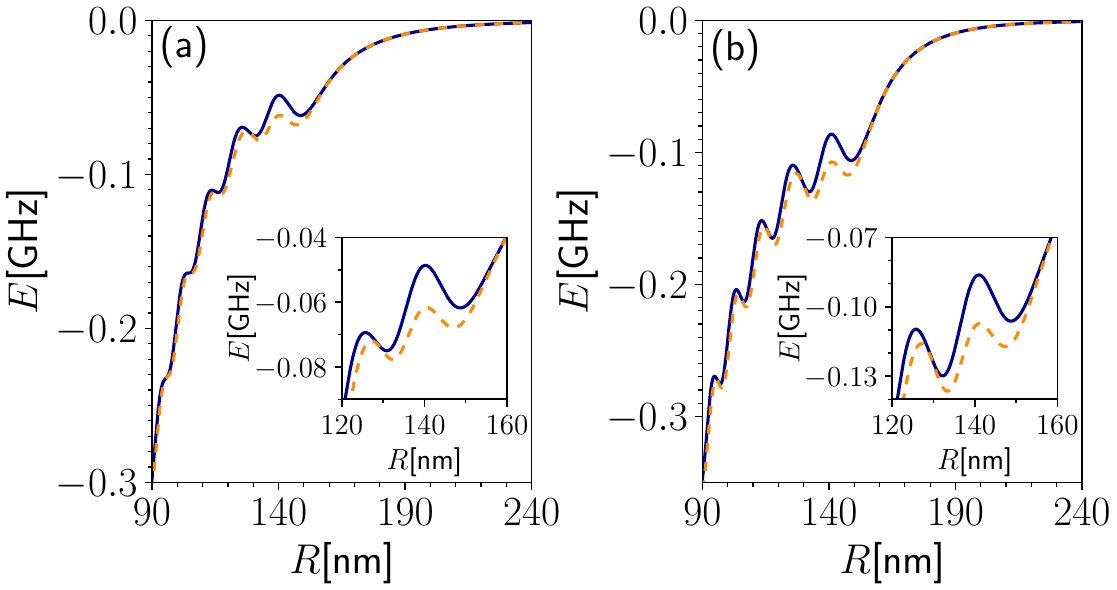}
 \caption{
Adiabatic electronic potential curves of the Cs-RbCs Rydberg molecule with $M_J=0$. 
(a) for Cs$(41p)$-RbCs($N=0$), (b) for Cs$(40d)$-RbCs($N=0$). Blue solid lines include the electron scattering with RbCs and orange dashed lines represent the curves with this term neglected.
In panels (a) and (b), the zero energies are set to the \ry energies 
$E(\text{Cs}(41p))$ and $E(\text{Cs}(40d))$, respectively.}
 \label{fig:comparison}
\end{figure}

\begin{figure*}
 \includegraphics[width=0.94\linewidth]{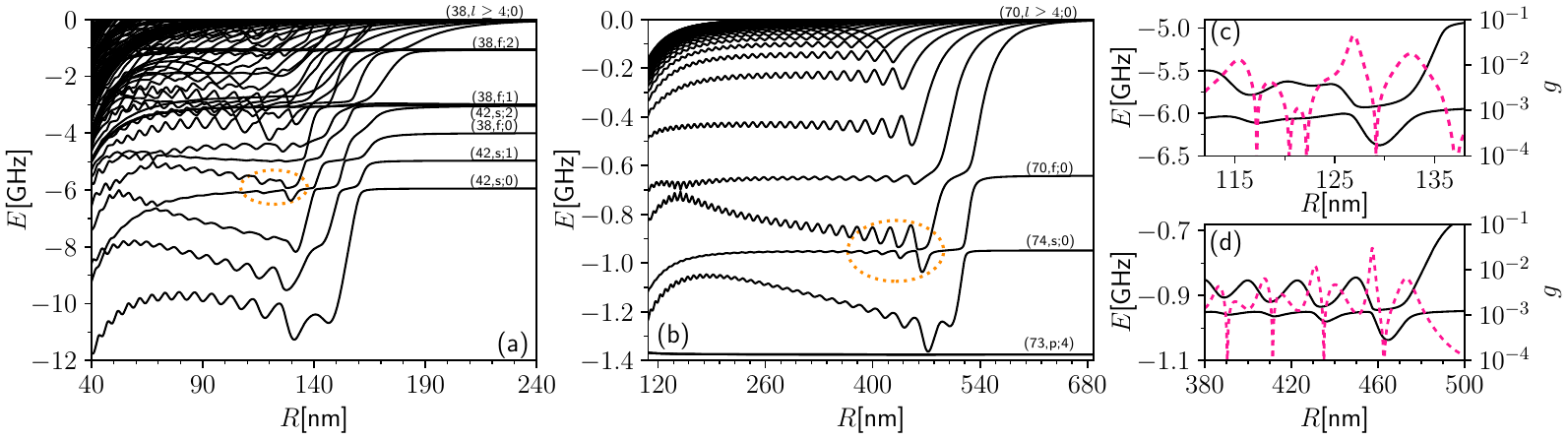}
 \caption{For the Cs-RbCs \ry molecule, adiabatic electronic potential energy curves for $M_J=0$
evolving from (a) the \ry states Cs$(n=38,l\ge4)$ and the RbCs in the ground state, 
Cs$(42s)$-RbCs($N\le 2$) and Cs$(38f)$-RbCs($N\le 2$) and 
(b) the \ry state Cs$(n=70,l\ge4)$ and the RbCs in the ground state, 
Cs$(70f)$-RbCs($N=0$), Cs$(74s)$-RbCs($N=0$) and Cs$(73p)$-RbCs($N=4$). 
The labels $(n,l;N)$ are used to identify the \ry potentials Cs$(nl)$-RbCs($N$). In panels
(c) and (d), the left axis shows potential curves (black solid lines) and the right axis shows the Landau-Zener coupling
$g=| \langle \Psi_i | \frac{\partial}{\partial R} |\Psi_j\rangle|$ (magenta dashed lines)
for the avoided crossings encircled in panels~(a) and (b), respectively.}
 \label{fig:apc_n_38}
\end{figure*}

We first quantify the effect of electron-molecule scattering on the adiabatic electronic potentials.
In the collision of an electron with RbCs(X$^1\Sigma$) negative ions could be formed, and the measured electron affinity of RbCs is $E_A=0.478 \pm 0.020$~eV~\cite{EATON1992141}.  {\it Ab initio} calculations of the low-energy electron-molecule scattering length are notoriously difficult. We therefore estimate the magnitude of the scattering length using $a_S=(2E_A)^{-1/2}\sim 5.34~a_0$. 
The existence of an $E_A$ for RbCs implies the sign of the scattering length is positive.
We use a $k$-independent scattering length in the $S$-wave Fermi-pseudopotential~\autoref{eq:Hamil_molecule}. 
The effect of this interaction is illustrated in~\autoref{fig:comparison} for the adiabatic potentials of
Cs($41p$)-RbCs($N=0$) and Cs($40d$)-RbCs($N=0$), which have been computed 
with (blue solid lines) and without (orange dashed lines) the pseudopotential in Eq.~\ref{eq:Hamil_molecule}.
Because $a_S >0$, the effect of the Fermi-pseudopotential is to {blue-shift the energies from those obtained with only the charge-dipole interactions in Eq.~\ref{eq:field_rydberg_e_core}}. We find, as expected, that the largest difference between these potentials is 
near the last maximum of the \ry wave function where we calculate a difference of around $48.6$~MHz between the two curves.

\subsection{Cs$^*$-RbCs Rydberg polyatomic molecules}

We consider two possible experimental protocols to create
 ultralong range polyatomic Rydberg molecules.
Firstly, the ground state Cs atom and the RbCs molecule are prepared in the motional ground state of two 
separate optical tweezers, 
as demonstrated for Rb in Ref.~\cite{guttridge2023}. Then, the optical tweezers are merged, ideally achieving a high 
occupation of the ground state of the relative motion of Cs and RbCs. Finally, the Cs atom is excited to a \ry state
so that RbCs lies within the electron orbit. 
Within the same optical tweezer, the spatial separation between the atom and molecule should be smaller 
might be smaller than 
$200$~nm, and the condition of lying within the orbit could be satisfied with a
two-photon excitation to the Cs($42s$) \ry state. 
An alternative route is to
leave the atom and molecule in their separate optical tweezers and to then excite the Cs atom.
In this case, the spatial separation between the centers of the tweezers of $\lambda/2\sim 500$~nm is achievable, 
and a higher principal quantum number for the \ry excitation is required. In this work, we take Cs($74s$) 
as an example to illustrate
this second option.

The electronic structures of the TriMol Cs($42s$)-RbCs($N=0$) and Cs($74s$)-RbCs($N=0$) are presented in 
Figs.~\ref{fig:apc_n_38}~(a) and~(b), respectively. Due to the peculiar quantum defect of Cs($ns$), the 
adiabatic potential curve for the TriMol Cs($ns$)-RbCs($N=0$) lies among those electronic states evolving 
from the \ry degenerate manifold Cs($n-4$, $l\ge 4$) combined with RbCs($N=0$). As a consequence, the 
electronic spectrum of these TriMol Cs$^*$-RbCs is characterized by numerous avoided crossings, as observed 
in panels (a) and~(b) of~\autoref{fig:apc_n_38}. Furthermore, because of the small RbCs rotational constant, 
the adiabatic potentials of the electronic states Cs($42s$)-RbCs($N$) with $N\le 2$, also become 
immersed among those 
evolving from Cs($n=38$, $l\ge 4$)-RbCs($N=0$), creating a highly complex spectrum.

\begin{figure*}
 \includegraphics[width=0.98\linewidth]{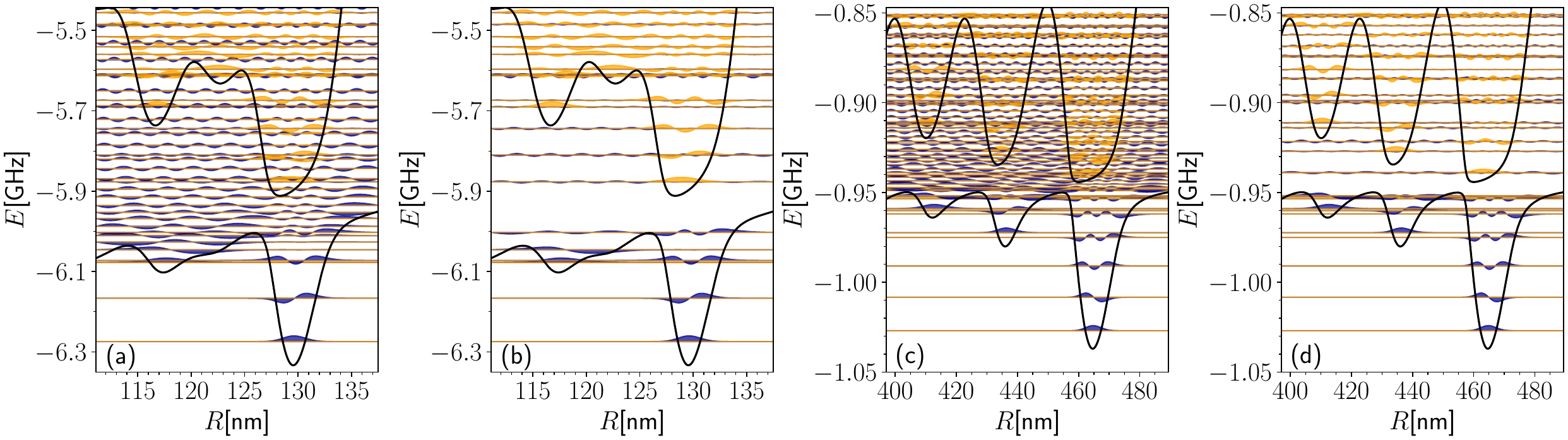}
 \caption{For the electronic states Cs($42s$)-RbCs($N=2$) and Cs($38f$)-RbCs($N=1$) of the TriMol 
encircled in~\autoref{fig:apc_n_38}~(a), eigenenergies and  their vibrational wave functions obtained from (a)
the coupled Schr\"odinger equation~\eqref{eq:matrix_hamil} and (b) after stabilization technique is applied. 
Panels (c) and (d) show similar results for the lowest Cs($n=70, l> 3$)-RbCs($N=0$) and 
Cs($70f$)-RbCs($N=0$) potentials encircled in~\autoref{fig:apc_n_38}~(b).
The vibrational wave functions are shifted from their
corresponding energies and rescaled for better visibility, and  the components $\chi^d(R)$ and $\chi^u(R)$ are plotted in blue and yellow, respectively.}
 \label{fig:non-adiabatic_WF_n_31_50_51}
\end{figure*}

The experimental observation of a TriMol requires the formation of a bound vibrational state in one of these
electronic potentials in~\autoref{fig:apc_n_38}, which 
would be heralded by the appearance of a redshidted feature in the Cs Rydberg spectrum.
In the avoided crossings marked in Figs.~\ref{fig:apc_n_38}~(a) and~(b) the coupling between neighboring electronic states is not negligible
and provokes the predissociation of the bound vibrational states. 
In these regions the Born-Oppenheimer 
approximation breaks down, and the total wave function can be expressed as 
\begin{equation}
\label{eq:wf_expansion}
\Phi({\bf r},\!\Omega,\!R)=\sum_{i=d,u}\Psi_i({\bf r},\!\Omega;\!R)\frac{\chi^i(R)}{R}, 
\end{equation}
where $\Psi_i({\bf r},\!\Omega;\!R)$ is the eigenfunction of the Halmiltonian~\eqref{eq:Hamil_adiabatic} and is 
given by~\autoref{eq:basis_expansion}, 
and the index $i=d,u$ indicates the potentials down and up, respectively.
The reduced vibrational wave functions $\chi^i(R)$ satisfy the following coupled Schr\"odinger 
equation~\cite{horstbook}
\begin{equation}
\label{eq:matrix_hamil}
\begin{pmatrix}
T+V_d+{\mathcal{A}}_{dd} & {\mathcal{A}}_{du} \\
{\mathcal{A}}_{ud} & T+V_u+{\mathcal{A}}_{uu}
\end{pmatrix}
\begin{pmatrix}
\chi^d \\
\chi^u
\end{pmatrix}=E
\begin{pmatrix}
\chi^d \\
\chi^u
\end{pmatrix},
\end{equation}
with $T=-\frac{\hbar^2}{2m}\frac{d^2}{dR^2}$ the reduced kinetic energy and 
 $m$ the reduced mass of the \ry TriMol.
 The coupling terms are 
 $\mathcal{A}_{ij}=\langle\Psi_i | T|\Psi_j \rangle-\frac{\hbar^2}{m}\langle \Psi_i|\frac{d}{dR}|\Psi_j\rangle\frac{d}{dR}$, 
 whereas the potentials 
$V_i=\langle \Psi_i | H | \Psi_i \rangle $ are the eigenvalues of the adiabatic Hamiltonian~\eqref{eq:Hamil_adiabatic},
with $i,j=d,u$.
Note that equation~\eqref{eq:matrix_hamil} is only valid for bound states of the \ry molecule with zero rotational angular momentum.

The states of the coupled potential energy curves Cs($42s$)-RbCs($N=2$) and Cs($38f$)-RbCs($N=1$)
computed within {the non-adiabatic Schr\"odinger equation~\eqref{eq:matrix_hamil}}, 
are presented in~\autoref{fig:non-adiabatic_WF_n_31_50_51}~(a)
together their wave functions.  
The contributions to the wave functions from the potentials down and up, \ie, $\chi^d(R)$ and  $\chi^u(R)$, respectively,
are plotted in different colors.
To determine which states are truly bound, we solve the coupled Schr\"odinger~\autoref{eq:matrix_hamil} varying the lower limit of the radial coordinate
and use the stabilization method~\cite{PhysRevA.1.1109}. 
The bound states are those levels whose energy could be considered independent of $R_{min}$, 
and they are presented in panel (b) of~\autoref{fig:non-adiabatic_WF_n_31_50_51}. 
For the sake of completeness, the corresponding stabilization diagram of the energies in~\autoref{fig:non-adiabatic_WF_n_31_50_51}~(a) is shown 
in~\autoref{fig:stabilization} of the appendix.  
A similar study is done for the electronic potential evolving from Cs($70f$)-RbCs($N=0$) taking into account  non-adiabatic coupling
to the upper potential evolving from Cs($70,l\ge4$)-RbCs($N=0$), 
all eigenstates for a fixed radial range are presented 
in~\autoref{fig:non-adiabatic_WF_n_31_50_51}~(c). Whereas the truly bound ones, obtained  after the stabilization analysis is applied, 
are shown in~\autoref{fig:non-adiabatic_WF_n_31_50_51}~(d).  

\begin{figure}
   \includegraphics[width=0.98\linewidth]{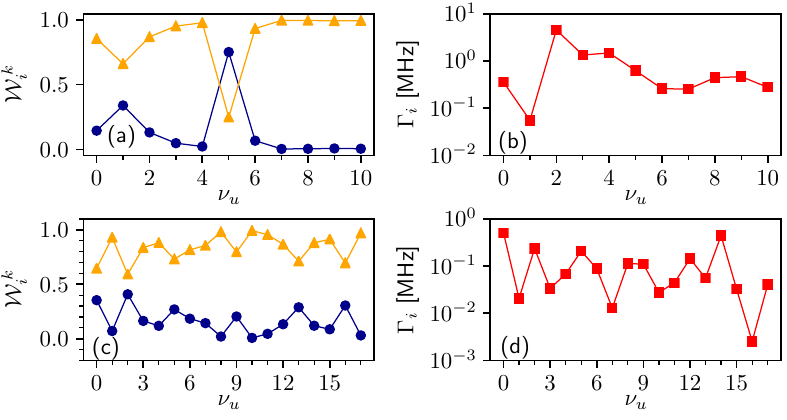}
 \caption{ 
For the  vibrational states in the upper potentials in Figs.~\ref{fig:non-adiabatic_WF_n_31_50_51} (b) and 
(d), panels (a) and (c) present weights, as defined in~\autoref{eq:weight_coupled}, of the vibrational wave 
function in the lower and upper potentials  
\ie,  ${\cal W}_i^d$ (blue circles) and ${\cal W}_i^u$ (yellow triangles),
as a function of the vibrational quantum number in these upper potentials; 
(b) and (d) present decay rates as defined in~\autoref{eq:decay_rates} as a function of the vibrational quantum number in these upper 
potentials, respectively.  
} 
 \label{fig:decay_rates_38}
\end{figure}
The broad avoided crossings hardly influence the deeply bound states in the outer well potential, and
their wave functions are dominated by the components in the lower potential, \ie, ${\cal W}_i^d\sim 1$, 
and see~\autoref{eq:weight_coupled} for the weights,
as the contribution 
from the upper potential is very small.
In contrast, the impact is significant for the states lying on the upper potential.
Their wave functions are dominated by the contribution from this upper potential, 
but still possess a significant weight from lower one as shown in~\autoref{fig:decay_rates_38}~(a) and~(c).
The three lowest states within the right well of Cs($38f$)-RbCs($N=1$) possess the largest mixing, together 
with the $\nu_{u}=5$ state, which due to the spatial extension of its wave function 
shows the highest coupling to the lower potential.
Note that for the rightmost well in the upper potential in ~\autoref{fig:non-adiabatic_WF_n_31_50_51}~(c),
which evolves from Cs($70,l\ge4$)-RbCs($N=0$), 
we encounter three states associated with the first excited vibrational state, \ie,
the part of their wave functions on the upper potential possess a node. 
For two of the states, the contributions from the lower potential to their total wave functions, see~\autoref{eq:weight_coupled}, are larger than $73\%$, whereas for
third state, it is $47\%$. However, they are not bounded because their energies are not stable as the lower limit 
of the radial coordinate is changed when solving the coupled Schr\"odinger equation in the stabilization diagram. 

To illustrate the coupling induced at the avoided crossings,
we have computed the nonadiabatic decay rates~\autoref{eq:decay_rates} of the bound states in the upper potential  to the continuum states. 
The results for the vibrational states within the Cs($38f$)-RbCs($N=1$) and 
the lowest electronic state from the degenerate manifold Cs($70,l\ge4$)-RbCs($N=0$)
are shown in Figs.~\ref{fig:decay_rates_38}~(b) and~(d), respectively.
The trend in the decay rate for the vibrational states in Fig.~\ref{fig:non-adiabatic_WF_n_31_50_51}~(b) is 
as expected; the higher vibrational states away from the avoided crossing region decay more slowly than 
the states near the crossing. 
For those states located in a potential well, the nodal structure of the vibrational wave function 
gives rise to the oscillatory behaviour of the decay rate.
For the vibrational states in the upper potential in Fig.~\ref{fig:non-adiabatic_WF_n_31_50_51}~(d), the oscillation in the decay rate is a reflection of the interference of the wave function spread in the three wells.

\begin{figure}
 \includegraphics[width=0.97\linewidth]{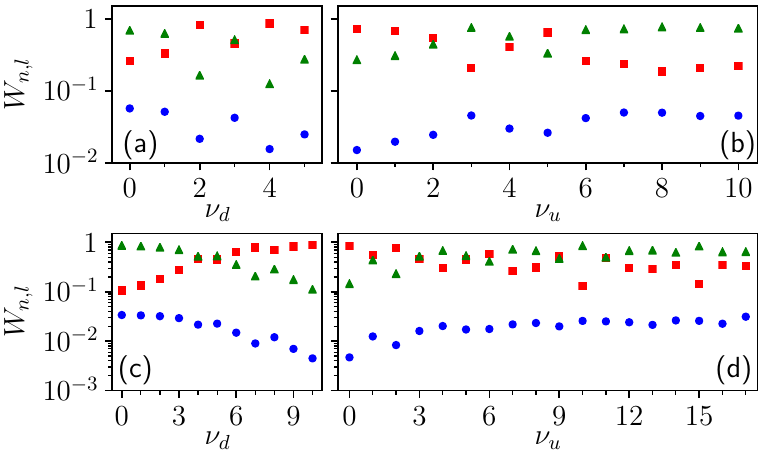}
 \caption{Integrated weights of the \ry $l$-partial waves in the electronic wave function, $W_{n,l=0}$ (red squares), 
 $W_{n,l=3}$ (blue dots) and $\sum_{l\ne 0,3}W_{n,l}$ (green triangles), see~\autoref{eq:weight_l_nu},
 for the bound states within the lower and upper potential energy curves (a) and (b), respectively,
 in~\autoref{fig:non-adiabatic_WF_n_31_50_51}~(b). Panels  
 (c) and (d) show analogous results for the bound states in the lower and upper potentials
 in~\autoref{fig:non-adiabatic_WF_n_31_50_51}~(d).}
 \label{fig:weights_l_38}
\end{figure}

\begin{figure}
 \includegraphics[width=0.95\linewidth]{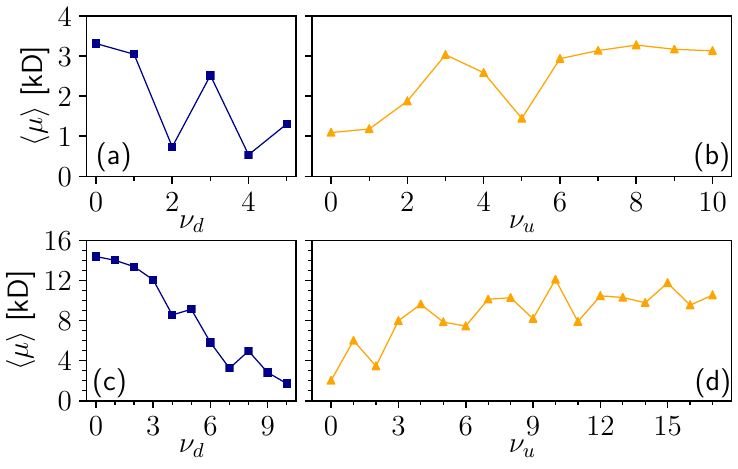}
 \caption{For the bound states within the lower and upper potentials  
in~\autoref{fig:non-adiabatic_WF_n_31_50_51}~(b), panels (a) and (b) show the 
 permanent electric dipole moment defined in~\autoref{eq:expectation_values_integrated_l},  respectively.
Panels (c) and (d) show analogous results for the bound states in the lower and upper potentials 
 in~\autoref{fig:non-adiabatic_WF_n_31_50_51}~(d). The sudden decrease in the dipole moment for the vibrational quantum numbers $\nu_{d}= 2$ and $4$ in panel (a) is 
 because these states are localized in the inner well of the down potential and, therefore, being less affected 
 by the avoided crossing in~\autoref{fig:non-adiabatic_WF_n_31_50_51}~(b).
}
 \label{fig:weights_38}
\end{figure}

For the vibrational bound states within the lower and upper potentials 
in~\autoref{fig:non-adiabatic_WF_n_31_50_51}~(b) and~(d), we have also analyzed the hybridization of the 
\ry partial waves due to the charge-dipole interaction. 
The weights of the $s$ and $f$ partial waves, and the sum of the rest, 
see~\autoref{eq:weight_nlN_coupled}, 
to the wave functions of the vibrational bound states of the Cs($42s$)-RbCs($N=2$) and Cs($38f$)-RbCs($N=1$) potentials
are shown in~\autoref{fig:weights_l_38}~(a) and~(b), respectively.
The $s$-wave is dominant only for those bound states close to the avoided crossing region, \ie,
the high (low)-lying states in the lower (upper) potential. This is due to the avoided crossing with the 
potentials Cs($42s$)-RbCs($N=0$) and Cs($42s$)-RbCs($N=1$) at larger internuclear separations. 
For all states, the \ry partial waves with $l\ge 3$ possess significant weights, having the largest 
contribution for excited states on the upper potentials.
These results indicate a significant hybridization of the \ry partial waves, giving rise to huge permanent electric dipole moments ($\sim$kD),
as illustrated in~\autoref{fig:weights_38}~(a) and~(b).

Figures~\ref{fig:weights_l_38}~(c) and~(d) present the \ry partial waves contributions to 
the vibrational states in the Cs($70f$)-RbCs($N=0$) and Cs($70,l\ge4$)-RbCs($N=0$) potentials, respectively. 
We encounter a strong mixing, indeed, they do not inherit a dominant $f$-wave 
character from the potential correlating to the Cs($70f$)-RbCs($N=0$) dissociation limit.
Again, the contribution of the $s$-partial wave is dominant for those states close 
to the avoided crossing region, \ie, the highest- and lowest-lying levels in the lower and upper potentials 
in~\autoref{fig:non-adiabatic_WF_n_31_50_51}~(d), 
respectively. As a consequence, they all possess {massive} electric dipole moments as
presented in \autoref{fig:weights_38}~(c) and~(d).

\begin{figure}
 \includegraphics[width=0.98\linewidth]{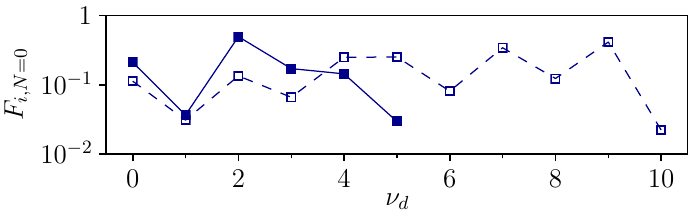}
 \caption{Weighted Franck-Condon factors  with $l=0$ \ry partial wave and $N=0$, assuming the Rabi frequency for 
the corresponding two-photon Rydberg excitation $\Omega_{ns}=1$,
 see~\autoref{eq:FCF}, to the
 bound states within the lower potential energy curves 
 Cs($42s$)-RbCs($N = 2$) in~\autoref{fig:non-adiabatic_WF_n_31_50_51}~(b) (filled squares),
 and Cs($70f$)-RbCs($N =0 $) in~\autoref{fig:non-adiabatic_WF_n_31_50_51}~(d) (empty squares).}
 \label{fig:FC_38_70}
\end{figure}

For the vibrational bound states within the lower potential energy curve, we have computed the weighted 
Franck-Condon factors in~\autoref{fig:FC_38_70}, taking the Rabi frequency for the two-photon Rydberg excitation 
to the $ns$ \ry states as $\Omega_{ns}=1$.
The Franck-Condon factors strongly depend on the nodal structure of the vibrational wave function. 
For the Cs($42s$)-RbCs($N=2$) potential, the 
lowest lying state of the left well possesses the largest Franck-Condon factor, which is also due to its
significant $s$-wave and $N=0$ contributions to the wave function. 
For the Cs($70f$)-RbCs($N =0 $) potential, we encounter that excited states 
possess larger Franck-Condon factors, for instance, the lowest
lying states within the leftmost potential well, and the second excited state in the central well. 
This could be due to the larger contribution of the \ry $s$ partial wave to their wave function.
These results indicate that in both cases, there exits several vibrarional bound states that could be used
for the photoassociation of these TriMols.

\section{Concluding remarks and outlook}
\label{sec:conclusions}
We have investigated triatomic ultralong-range Rydberg molecules formed by a Rydberg cesium atom 
and a RbCs diatomic molecule.
Our theoretical description includes the charge-dipole interaction arising from the coupling of the 
permanent dipole moment of RbCs with the electric field produced by the \ry electron and core, 
and the $S$-wave scattering of the slow \ry electron from the molecule. 
Due to the peculiar quantum defects of Cs, the charge-dipole interaction induces 
a coupling between the $s$-wave quantum defect Rydberg states and the nearest degenerate 
hydrogenic manifold, as both states are combined with RbCs in its rotational ground state. 

Assuming that the \ry atom and the diatomic molecule are confined in optical 
tweezers~\cite{guttridge2023}, we focus on the \ry molecules evolving from Cs($42s$)-RbCs($N=0$) and Cs($74s$)-RbCs($N=0$), whose 
interspecies separations correspond to the two possible scenarios with Cs and RbCs being in the
same tweezer or in separate ones, respectively.
For these electronic states, we have explored the role played by the non-adiabatic coupling near the avoided crossings,
their main features and vibrational structure. In both cases, we have identified vibrational
bound states with large Franck-Condon factors, that could be used to form the polyatomic \ry molecules. 

Near resonant coupling between \ry states and molecular rotation induces long-range dipole-dipole interactions. Charge-dipole coupling induces gigantic shifts and permanent polyatomic molecular dipole moments. These state-specific interactions can be utilized to perform conditional and nondestructive readout of molecular state~\cite{Kuznetsova2011,Kuznetsova2016,Gawlas2020,Patsch2022,Wang2022,Zhang2022}.
In addition, quantum gates between molecules could be implemented using 
the \ry  states of Cs as facilitators~\cite{Wang2022,Zhang2022}.
A number of different ultracold bialkali molecules containing Cs atoms have been successfully produced in a variety of settings. For instance, production of molecules in the absolute ground state has been achieved for LiCs using photoassociation~\cite{Deiglmayr2008}, NaCs using magnetoassociation in optical tweezers~\cite{Cairncross2021} and bulk gases~\cite{Stevenson2022}, and RbCs molecules have been produced in bulk gases~\cite{Molony2014,Takekoshi2014} and optical tweezers~\cite{guttridge2023}. In addition, there are good prospects for reaching the ground state of ultracold KCs~\cite{Groebner2016}. For all these polar alkali dimers, we present in~\autoref{tb:transitions_energy} 
a selection of the pairs of rotational levels and of \ry states in Cs,
that could be used for this near-resonant interaction, along with the energy difference between these transitions. 
These energy splittings could be brought closer to resonance by  applying electric or magnetic fields to smoothly tune the \ry states.

{The precision tunability of \ry atoms and the stability of molecular dipoles combine to induce anisotropic long-range interactions with new scales and functionality. This hybrid atom-molecule system holds potential for investigation of an entire class of quantum processes, including quantum magnetization and quantum simulations.}
\begin{table}
\caption{Transitions of the Cs atom and diatomic molecules formed from Cs and energy differences of these
transitions that could be used to create quantum gates.}
\begin{center}
\begin{tabular}{|c|c|c|}
 \hline
 Cesium & Molecule and & Energy difference\\ 
\ry Transition & Rotational Transition & [MHz]\\ 
 \hline
 $95S_{1/2}\, \to \, 92P_{3/2}$ &LiCs, $N=1\to2$ & $2.3$\\
 $47D_{5/2}\, \to \, 48P_{3/2}$ & NaCs, $N=1\to2$ & $4.3$\\
$118D_{3/2}\, \to \,115P_{1/2}$ & KCs, $N=0\to1$ & $0.4$\\
 $85D_{3/2}\, \to \, 86P_{3/2}$ & RbCs, $N=0\to1$ &$0.6$\\
 \hline
\end{tabular}
\end{center}
\label{tb:transitions_energy}
\end{table}%

 \begin{acknowledgments} 
D.M.A. and R.G.F. gratefully acknowledge financial support by the Spanish projects PID2020-113390GB-I00 (MICIN),
PY20$\_$00082 (Junta de Andalucía), and A-FQM-52-UGR20 (ERDF-University of Granada) 
and the Andalusian research group FQM-207.
A.G and S.L.C. acknowledge support from the UK Engineering and Physical Sciences Research Council (EPSRC) Grants EP/P01058X/1, EP/V047302/1, and EP/W00299X/1, UK Research and Innovation (UKRI) Frontier Research Grant EP/X023354/1, the Royal Society, and Durham University.
H.R.S. acknowledges support at ITAMP through a grant by the NSF. 

\end{acknowledgments}

\appendix

\section{Expressions}
\label{appen:expressions}

In this appendix, we describe the quantities analyzed to get a better physical insight of the features of the TriMol.
To illustrate the induced coupling due to the charge-dipole interaction, we compute
 the weight of a certain partial wave of the \ry electron in a $k$-electronic state of the TriMol wave
 function~\ref{eq:basis_expansion}, which is given by
\begin{equation}
{\cal C}_{n,l}(R)=\sum_{N,J}|C_{n,l,N}^{J}(R)|^2\, ,
\label{eq:weight_l}
\end{equation}
where the sum in $J$ is for $|l-N|\le J\le l+N$, 
and satisfies $\sum_{n,l}{\cal C}_{n,l}(R)=1$. 
For a vibrational bound state, we define
the weight of this Rydberg-electron partial wave as 
\begin{equation}
\label{eq:weight_l_nu}
W_{n,l}^\nu=
\int(\chi(R))^{*}{\cal C}_{n,l}(R) \chi(R) dR, \,\,
\end{equation}
and $\sum_{n,l}W_{n,l}^k=1$. 
Note that $\chi(R)$ represents the reduced wave function of a vibrational state,
and satisfies the following Schr\"odinger equation
\begin{equation}
\label{eq:rovibrational}
\Big[-\frac{\hbar^2}{2m}\frac{d^2}{dR^2}+ V(R)\Big]\chi(R)=E\chi(R)
\end{equation}
with $m$ being the reduced mass of the TriMol, and $V(R)$ the corresponding adiabatic potential energy curve.

We also estimate the electric dipole moment of this vibrational state by 
\begin{equation}
\langle\mu\rangle=
\int (\chi(R))^*D_{ryd}(R) \chi(R) dR\, ,
\label{eq:expectation_values_integrated_l}
\end{equation}
with the $R$-dependent electric dipole moment given by
\begin{equation*}
D_{ryd}(R)=
e\int \Psi^*(\mathbf{r},\Omega;R)
r\cos \theta_e
\Psi(\mathbf{r},\Omega;R)d^3r d\Omega\, ,
\label{eq:expec_dipole_moment}
\end{equation*}
where $\theta_e$ is the polar angle of the \ry electron. This integral is non-zero if there are 
partial waves with different parity, \ie, $\Delta l=\pm 1$, 
contributing to the wave function~\ref{eq:basis_expansion}.

For a bound state in the coupled electronic potentials, we define the weight of the wave function in the potentials
down and 
up as 
\begin{equation}
{\cal W}_i^k=\int |\chi^k_i(R)|^2dR
\label{eq:weight_coupled}
\end{equation}
where $\chi^k_i(R)$ represents the part of reduced wave function in the $k$-potential with
$k=d,u$, and $i$ identifies the vibrational states. For all bound states, it holds ${\cal W}^d_i+{\cal W}^u_i=1$. 
For each of these bound states, the weights of the Rydberg-electron wave function 
are defined, respectively,  as
\begin{equation}
W_{n,l}=\sum_{k=d,u}W_{n,l}^k\,.\quad 
\label{eq:weight_nlN_coupled}
\end{equation}

The non-adiabatic decay rates due to coupling near the avoided crossings care expressed as~\cite{Duspayev_2022}
\begin{equation}
\Gamma_i=
\frac{2\pi}{\hbar}|\langle \chi_i^d| A_{du}| \chi_j^u \rangle|^2
\label{eq:decay_rates}
\end{equation}
where $\mathcal{A}_{du}=\langle\Psi_d | T|\Psi_u \rangle-\frac{\hbar^2}{m}\langle \Psi_d|\frac{d}{dR}|\Psi_u\rangle\frac{d}{dR}$,
with $\chi_j^u$ is the vibrational wave function of a bound state in the upper potential, and 
$\chi_i^d$, is the the continuum wave function in the lower potential at the same energy. ~
Note that the scattering state $\chi_i^d$ is energy normalized, whereas computationally it is $L^2$ normalized. A numerical
way to obtain these energy normalized wave functions is described in Ref.~\cite{Gonzalez2007}.

In the coupled picture, the electronic dipole moment of a bound state is 
\begin{equation}
\langle i| \mu| i \rangle =
\int (\chi_i^{k_1}(R))^*D_{ryd}^{k_1,k_1}(R) \chi_i^{k_1}(R)dR\, ,
\label{eq:dipole_transition_ij}
\end{equation}
with the dipole moment of the electronic state $k_1$, with $k_1=d,u$, given by 
\begin{equation*}
D_{ryd}^{k_1,k_1}(R)=
e\int \Psi_{k_1}^*(\mathbf{r},\Omega;R)
r\cos \theta_e
\Psi_{k_1}(\mathbf{r},\Omega;R)d^3r d\Omega\,.
\end{equation*}

To estimate the probability of forming these \ry molecules by a two photon transition, we compute the 
 Franck-Condon factors weighted with $l=0$ \ry partial wave,
\begin{equation}
F_{i,N}=\sum_{k=d,u}\int(\chi_{i}^k(R))^{*} \Omega_{ns} C^{J,k}_{n,l=0,N}(R) \psi_{scat}(R) RdR
\label{eq:FCF}
\end{equation}
with $\psi_{scat}(R)$ being the scattering wave function of the initial open channel and $\chi_i^k(R)$ the
vibrational wave function of the final potential energy curve of the TriMol, and
$ \Omega_{ns} $ the Rabi frequency for a two-photon Rydberg excitation to the $ns$ state, which was taken equal to $1$ in our
calculations. We are assuming that the process is dominated by a single $ns$ \ry state, \ie, there is no 
significant admixture of different principle quantum numbers, then we can drop the sum over $n$.

\begin{figure}[t]
 \includegraphics[width=0.9\linewidth]{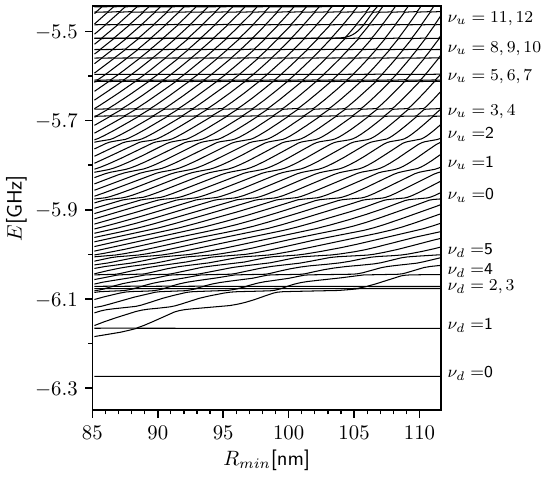}
\caption{For the two bound states of the 
electronics states of Cs-RbCs \ry molecule 
Cs($42s$)-RbCs($N=2$) and 
Cs($38f$)-RbCs($N=1$) presented in~\autoref{fig:non-adiabatic_WF_n_31_50_51}~(a), 
 stabilization diagram of the energies as a function of the lower minimum value of the radial coordinate $R_{min}$ used
 in the numerical method to solve the coupled Schr\"odinger equation~\eqref{eq:matrix_hamil}.
 The vibrational quantum number $\nu_d$  of the bound states is indicated in the right axis.}
\label{fig:stabilization}
\end{figure}

In~\autoref{fig:stabilization}, we present the stabilization
diagram of the electronic potentials of Cs-RbCs
shown in~\autoref{fig:non-adiabatic_WF_n_31_50_51}~(a). 
We have solved the Schr\"odinger equation~\eqref{eq:matrix_hamil} by using different discretization boxes 
for $R$ and modifying their lower limit $R_{min}$~\cite{Gonzalez2007}.
Those level whose energy could be considered independent of $R_{min}$
in~\autoref{fig:stabilization} are the states that are truly bound within these
potentials, and are shown in panel (b) of~\autoref{fig:non-adiabatic_WF_n_31_50_51}.

\bibliographystyle{apsrev4-2} 
\bibliography{TriMol}
\end{document}